\documentclass{article}
\usepackage{ismir,amsmath,cite,url}
\usepackage{graphicx}
\usepackage{color}
\usepackage{tabularx}
\usepackage{subcaption}
\usepackage{soul}
\usepackage{boldline}
\usepackage{multirow}

\title{Frame-level Instrument Recognition by Timbre and Pitch}

\oneauthor
 {Yun-Ning Hung and Yi-Hsuan Yang}
 {Research Center for IT Innovation, Academia Sinica, Taipei, Taiwan\\{\tt \{biboamy,yang\}@citi.sinica.edu.tw}}

\begin{document}

\maketitle
\begin{abstract}
Instrument recognition is a fundamental task in music information retrieval, yet little has been done to predict the presence of instruments in multi-instrument music for each time frame. This task is important for not only automatic transcription but also many retrieval problems. In this paper, we use the newly released MusicNet dataset to study this front, by building and evaluating a  convolutional neural network for making frame-level instrument prediction. We consider it as a multi-label classification problem for each frame and use frame-level annotations as the supervisory signal in training the network. 
%
%
Moreover, we experiment with different ways to incorporate pitch information to our model, with the premise that doing so informs the model the notes that are active per frame, and also encourages the model to learn relative rates of energy buildup in the harmonic partials of different instruments. Experiments show salient performance improvement over baseline methods. We also report an analysis probing how pitch information helps the instrument prediction task.
Code and experiment details can be found at \url{https://biboamy.github.io/instrument-recognition/}. 

\end{abstract}
\section{Introduction}
\label{sec:introduction}

Progress in pattern recognition problems usually depends highly on the availability of high-quality labeled data for model training. For example, in computer vision, 
the release of the ImageNet dataset \cite{imagenet}, along with advances in algorithms for training deep neural networks \cite{lecun2015deep}, has fueled significant progress in \emph{image-level} object recognition. 
The subsequent availability of other datasets, such as the COCO dataset \cite{coco}, provide bounding boxes or even \emph{pixel-level} 
annotations of objects that appear in an image, facilitating research on localizing objects in an image, semantic segmentation, and instance segmentation \cite{coco}.
Such a move from image-level to pixel-level prediction opens up many new exciting applications in computer vision \cite{garciaOO17arxiv}. 

Analogously, for many music-related applications, it is desirable to have not only \emph{clip-level} but also \emph{frame-level} predictions.
For example, expert users such as music composers may want to search for music with certain attributes and require a system to return not only a list of songs but also indicate the time intervals of the songs that have those attributes \cite{andersen16ismir}.
Frame-level predictions of music tags can be used for visualization and music understanding \cite{wang14icassp,JYnet}.
In automatic music transcription, we want to know the musical notes that are active per frame as well as figure out the instrument that plays each note 
\cite{duan14taslp}.
Vocal detection \cite{schluer16ismir} and guitar solo detection \cite{pati17aes} are another two examples that requires frame-level predictions.



Many of the aforementioned applications are related to the classification of sound sources, or instrument classification. However, as labeling the presence of instruments in multi-instrument music for each time frame is labor-intensive and time-consuming, most existing work on instrument classification uses either datasets of solo instrument recordings (e.g., the ParisTech dataset \cite{joder09taslp}), or datasets with only clip- or excerpt-level annotations (e.g., the IRMAS dataset \cite{bosch12ismir}).
While it is still possible to train a model that performs frame-level instrument prediction from these datasets, it is difficult to evaluate the result due to the absence of frame-level annotations.\footnote{Moreover, these datasets may not provide high-quality labeled data for frame-level instrument prediction. To name a few reasons: the ParisTech dataset \cite{joder09taslp} contains only instrument solos and therefore misses the complexity seen in multi-instrument music; the IRMAS dataset \cite{bosch12ismir} labels only the ``predominant'' instrument(s) rather than all the active instruments in each excerpt; moreover, an instrument may not be always active throughout an excerpt.}
As a result, to date little work has been done to specifically study frame-level instrument recognition, to the best of our knowledge (see Section \ref{sec:related_work} for a brief literature survey). 



The goal of this paper is to present such a study, by taking advantage of a recently released dataset called MusicNet \cite{thickstun2017learning}. 
The dataset contains 330 freely-licensed classical music recordings by 10 composers, written for 11 instruments, along with over 1 million annotated labels indicating the precise time of each note in every recording and the instrument that plays each note. 
Using the \emph{pitch} labels available in this dataset, Thickstun \emph{et al.} \cite{thickstun2018invariances} built a convolutional neural network (CNN) model that establishes a new state-of-the-art in multi-pitch estimation.
We propose that the frame-level \emph{instrument} labels provided by the dataset also represent a valuable information source. And, we try to realize this potential by using the data to train and evaluate a frame-level instrument recognition model.

Specifically, we formulate the problem as a multi-label classification problem for each frame and use frame-level annotations as the supervisory signal in training a CNN model with three residual blocks \cite{he16cvpr}. The model learns to predict instruments from a spectral representation of audio signals provided by the constant-Q transform (CQT) 
(see Section \ref{subsec:network} for details). 
Moreover, as another technical contribution, we investigate several ways to incorporate pitch information to the instrument recognition model (Sections \ref{subsec:pitch}), with the premise that doing so informs the model the notes that are active per frame, and also encourages the model to learn the energy distribution of partials (i.e., fundamental frequency and overtones) of different instruments \cite{agostini03eurasip,essid06taslp,barbedo11taslp,harmonic}. 
We experiment with using either the ground truth pitch labels from MusicNet, 
or the pitch estimates provided by the CNN model of Thickstun \emph{et al}. \cite{thickstun2018invariances} (which is open-source).
Although the use of pitch features for music classification is not new, to our knowledge few attempts have been made to jointly consider timbre and pitch features in a deep neural network model.
We present in Section \ref{sec:result} the experimental results and analyze whether and how pitch-aware models outperform baseline models that take only CQT as the input. 



%
\section{Related work}
\label{sec:related_work}

A great many approaches have been proposed for (clip-level) instrument recognition. Traditional approaches used domain knowledge to engineer audio feature extraction algorithms and fed the features to classifiers such as support vector machine \cite{ISMIR05,ISMIR06}. For example, Diment \emph{et al.} \cite{MODGDF} combined Mel-frequency cepstral coefficients (MFCCs) and phase-related features and trained a 
Gaussian mixture model. Using the instrument solo recordings from the RWC dataset \cite{goto2002rwc}, they achieved 96.0\%, 84.9\%, 70.7\% accuracy in classifying 4, 9, 22 instruments, respectively.  
Yu \emph{et al.} \cite{sparsecoding} used sparse coding for feature extraction and support vector machine for classifier training, obtaining 96\% accuracy in 10-instrument classification for the solo recordings in the ParisTech dataset \cite{joder09taslp}. 
Recently, Yip and Bittner \cite{yip17ismir-lbd} made open-source a solo instrument classifier that uses MFCCs in tandem with random forests to achieve 96\% frame-level test accuracy in 18-instrument classification using solo recordings from the MedleyDB multi-track dataset \cite{MedleyDB}. Recognizing instruments in multi-instrument music has been proven more challenging. For example, Yu \emph{et al.} \cite{sparsecoding} achieved 66\% F-score in 11-instrument recognition using a subset of the IRMAS dataset \cite{bosch12ismir}.


Deep learning has been increasingly used in more recent work. 
Deep architectures can ``learn'' features by training the feature extraction module and the classification module in an end-to-end manner \cite{lecun2015deep}, thereby leading to better accuracy than traditional approaches.  For example, Li \emph{et al.} \cite{Inst15} showed that feeding raw audio waveforms to a CNN achieves 72\% (clip-level) F-micro score in discriminating 11 instruments in MedleyDB, which MFCCs and random forest only achieves 64\%. Han \emph{et al.} \cite{predoInst} trained a CNN to recognize predominant instrument in IRMAS and achieved 60\% F-micro, which is about 20\% higher than a non-deep learning baseline. Park \emph{et al.} \cite{MRP} combined multi-resolution recurrence plots and spectrogram with CNN to achieved 94\% accuracy in 20-instrument classification using the UIOWA solo instrument dataset \cite{UIOWA}.

Due to the lack of frame-level instrument labels in many existing datasets, little work has focused on frame-level instrument recognition. The work presented by 
Schl\"{u}ter for vocal detection \cite{schluer16ismir} and by Pati and Lerch for guitar solo detection \cite{pati17aes} are exceptions, but they each addressed one specific instrument, rather than general instruments.
Liu and Yang \cite{JYnet} proposed to use clip-level annotations in a weakly-supervised setting to make frame-level predictions, but the model is for general tags.
Moreover, due to the assumption that CNN can learn high-level features on its own, domain knowledge of music has not been much used in prior work on deep learning based instrument recognition, though there are some exceptions \cite{lostanlen16ismir,pons16cbmi}.


Our work differentiates itself from the prior arts in two aspects. First, we focus on frame-level instrument recognition. Second, we explicitly employ the result of multi-pitch estimation \cite{Rachael,thickstun2018invariances} as additional inputs to our CNN model, with a design that is motivated by the observation that instruments have different pitch range and have unique energy distributions in the partials \cite{harmonic}. 


\section{Dataset}
\label{sec:db}

\begin{table}
\centering
\begin{tabular}{|r|r|r|r|} \hline
Number of instru- & \multicolumn{2}{c|}{Number of clips}
 & Pitch est.\\
 \cline{2-3}
ments used & Train set & Test set & accuracy\\
 \hline\hline
0 &  3 & 0 & ---\\
1 & 172 & 5 & 62.9\%\\
2 & 33 & 1 & 56.2\% \\
3 & 95 & 4 & 60.5\% \\
4 & 15 & 0 & 56.6\%\\
6 & 2 & 0 & 49.6\% \\
\hline
\end{tabular}
\caption{The number of clips in the training and test sets of MusicNet \cite{thickstun2017learning}, divided according to the number of instruments used (among the seven instruments we consider in our experiment) per clip (e.g., a piano trio uses 3 instruments). We also show the average frame-level multi-pitch estimation accuracy (using mir\_eval \cite{mireval}) achieved by the CNN model proposed by Thickstun \emph{et al}. \cite{thickstun2018invariances}.}
\label{tab:dataset}
\end{table}

Training and evaluating a model for frame-level instrument recognition is possible due to the recent release of the MusicNet dataset \cite{thickstun2017learning}. It contains 330 freely-licensed music recordings by 10 composers with over 1 million annotated pitch and instrument labels on 34 hours of chamber music performances. Following \cite{thickstun2018invariances}, we use the pre-defined split of training and test sets, leading to 320 and 10 clips in the training and test sets, respectively. 
As there are only seven different instruments in the test set, we only consider the recognition of these seven instruments in our experiment. They are \emph{Piano, Violin, Viola, Cello, Clarinet, Bassoon} and \emph{Horn}. For the training set, we do not exclude the sounds from the instruments that are not on the list, but these instruments are not labeled. Different clips use different number of instruments. See Table \ref{tab:dataset} for some statistics.
For convenience, each clip is divided into 3-second segments. We use these segments as the input to our model. We zero-pad (i.e., adding silence) the last segment of each clip so that it is also 3 seconds. Due to space limit, for details we refer readers to the MusicNet website (check reference \cite{thickstun2017learning} for the URL) and also our project website (see the abstract for the URL).

We note that the MedleyDB dataset \cite{MedleyDB} can also be used for frame-level instrument recognition, but we choose MusicNet for two reasons. First, MusicNet is more than three times larger than MedleyDB in terms of the total duration of the clips. Second, MusicNet has pitch labels for each instrument, while MedleyDB only annotates the melody line. 
However, as MusicNet contains only classical music and MedleyDB has more Pop and Rock songs, the two datasets feature fairly different instruments and future work can be done to consider they both.

\begin{figure*}[!htb]
\begin{subfigure}[b]{0.25\textwidth} 
	\includegraphics[width=\textwidth]{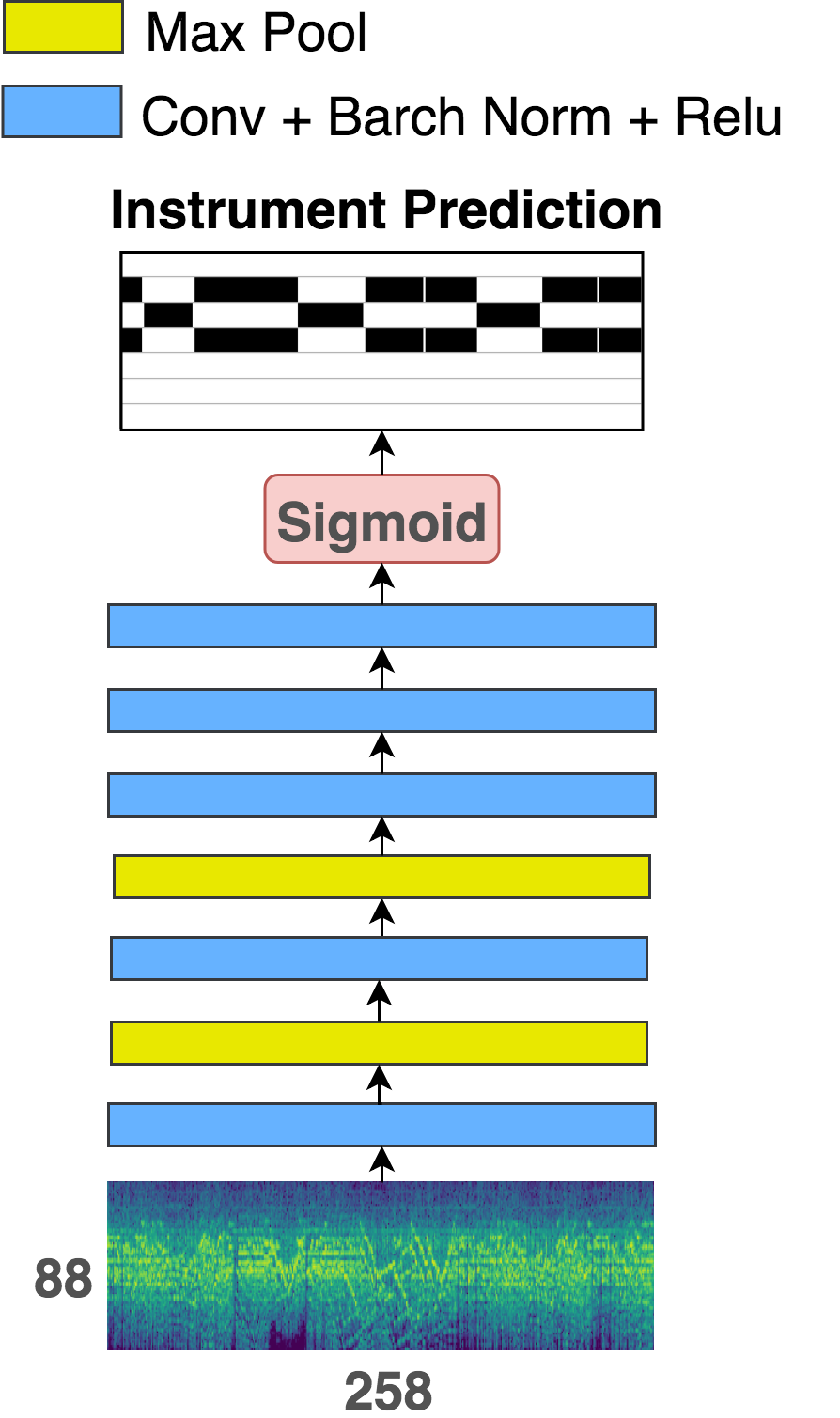}
    \caption{Baseline CNN \cite{JYnet}}
    \label{fig: baseline model}
\end{subfigure}
\hfill
\begin{subfigure}[b]{0.35\textwidth} 
	\includegraphics[width=\textwidth]{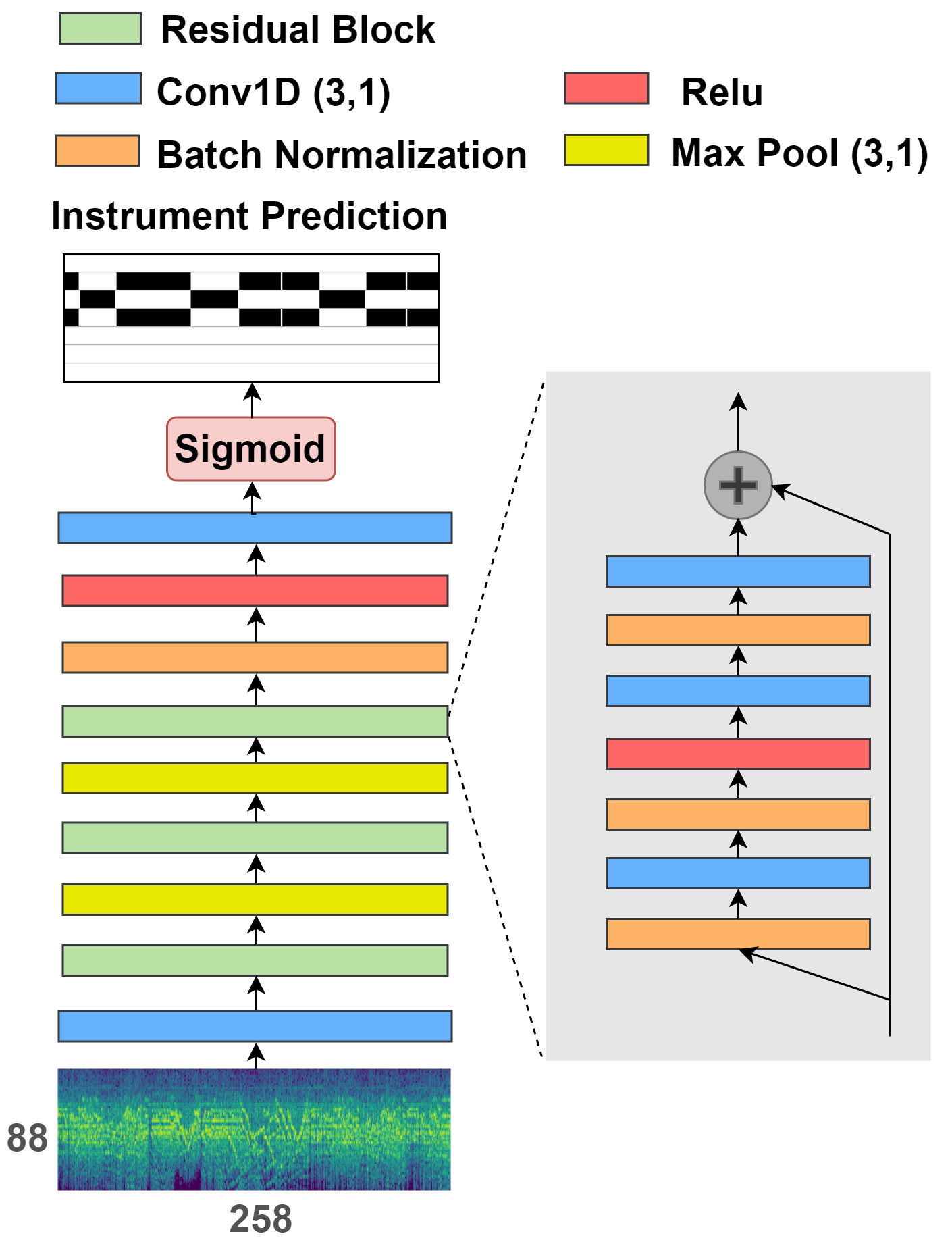}
    \caption{CNN + ResBlocks \cite{SYnet}}
    \label{fig: 1Dmodel}
\end{subfigure}
\hfill
\begin{subfigure}[b]{0.35\textwidth} 
	\includegraphics[width=\textwidth]{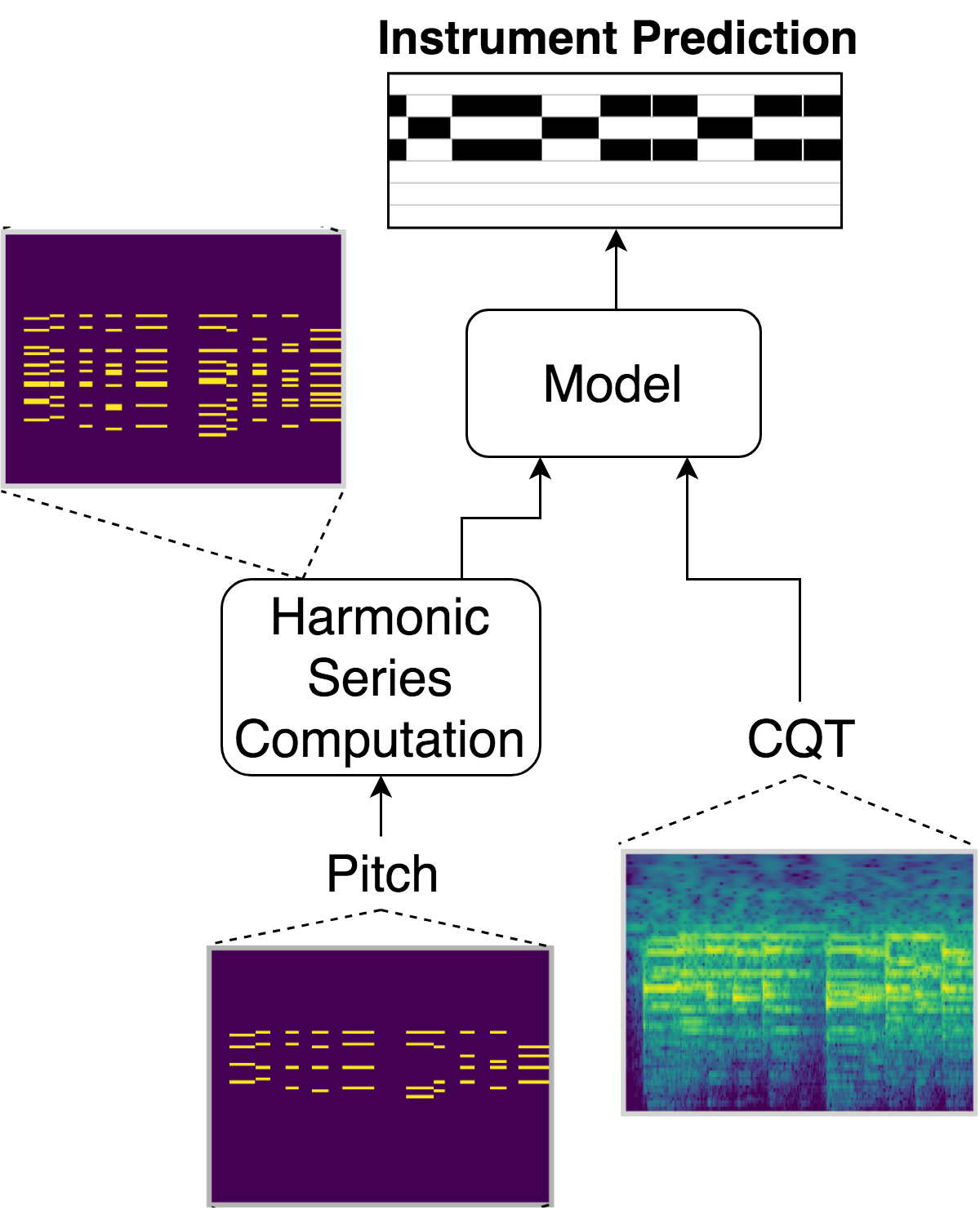}
    \caption{Pitch-aware model (CQT+HSF)}
    \label{fig: flow}
\end{subfigure}
\caption{Three kinds of model structure used in this instrument recognition experiment.}
\label{fig: model}
\end{figure*}


\section{Instrument Recognition Method}
\label{sec:method}

\subsection{Basic Network Architectures that Uses CQT}
\label{subsec:network}



To capture the timbral characteristics of each instrument, in our basic model we use CQT as the feature representation of music audio. CQT is a spectrographic representation that has a musically and perceptual motivated frequency scale \cite{cqt}. We compute CQT by \texttt{librosa} \cite{librosa}, with sampling rate 44,100 and 512-sample window size. 88 frequency notes are extracted with 12 bins per octave, which forms a matrix $\mathbf{X} \in \mathcal{R}^{258 \times 88}$ as the input data, for each inputting 3-second audio segment.

We experiment with two baseline models. 
The first one is adapted from the CNN model 
proposed by Liu and Yang \cite{JYnet}, which has been shown effective for music auto-tagging. Instead of using 6 feature maps as the input to the model as they did, we just use CQT as the input. Moreover, we use frame-level annotations as the supervisory signal in training the network, instead of training the model in a weakly-supervised fashion as they did. A batch normalization layer \cite{bn} is added after each convolution layer. Figure \ref{fig: baseline model} shows the model architecture. 

The second one is adapted from a more recent CNN model proposed by Chou \emph{et al.} \cite{SYnet}, which has been shown effective for large-scale sound event detection. Its design is special in two aspects. 
First, it uses 1D convolutions (along time) instead of 2D convolutions. While 2D convolutions  analyze the input data as a chunk and convolve on both spectral and temporal dimensions, the 1D convolutions (along time) might better capture frequency and timbral information in each time frame \cite{DCASE,SYnet}.
Second, it uses the so-called residual (Res) blocks \cite{he16cvpr,ResNet} to help the model learn deeper. Specifically, we employ three Res-blocks in between an \emph{early} convolutional layer and a \emph{late} convolutional layer. Each Res-block has three convolutional layers, so the network has a stack of 11 convolutional layers in total. 
We expect such a deep structure can learn well for a large-scale dataset such as MusicNet. Figure \ref{fig: 1Dmodel} shows its model architecture. 

 
%
%
%


\subsection{Adding Pitch}
\label{subsec:pitch}
Although people usually expect neural networks can learn high-level feature such as pitch, onset and melody, our pilot study shows that with the basic architecture the network still confuses some instruments (e.g., clarinet, bassoon and horn), and that onset frames for each instrument are not nicely located (see the second row of Figure \ref{fig:result}). 
We propose to remedy this with a pitch-aware model that explicitly takes pitch as input, in a hope that doing so can amplify onset and timbre information. We experiment with several methods for inviting pitch to join the model. 

\subsubsection{Source of Frame-level Pitch Labels}
\label{subsubsec:pitch_feature}

We consider two ways of getting pitch labels in our experiment. One is using human-labeled ground truth pitch labels provided by MusicNet. However, in real-word applications, it is hard to get 100\% correct pitch labels. Hence, we also use pitch estimation predicted by a state-of-the-art multi-pitch estimator proposed by Thickstun \emph{et al.} \cite{thickstun2018invariances}. The author proposed a translation-invariant network which combines traditional filterbank with a convolutional neural network. The model shares parameters in the log-frequency domain, which exploits the frequency invariance of music to reduce the number of model parameters and to avoid overfitting to the training data. The model reaches the top performance in the 2017 MIREX Multiple Fundamental Frequency Estimation evaluation \cite{MIREX}. The average pitch estimation accuracy, evaluated using mir\_eval \cite{mireval}, is shown in Table \ref{tab:dataset}.

\subsubsection{Harmonic Series Feature} 
\label{subsubsec:PHF}

Figure \ref{fig: flow} depicts the architecture of a proposed pitch-aware model. In this model, we aim to exploit the observation that the energy distribution of the partials constitutes a key factor in the perception of instrument timbre \cite{harmonic}. 
Being motivated by \cite{Rachael}, we propose the \emph{harmonic series feature} (HSF) to capture the harmonic structure of music notes, calculated as follows. 
We are given the input pitch estimate (or ground truth) 
$\mathbf{P}_0 \in \mathcal{R}^{258 \times 88}$, which is a matrix with the same size as the CQT matrix.
The entries in $\mathbf{P}_0$ take the value of either 0 or 1 in the case of ground truth pitch labels, and the value in $[0,1]$ in the case of estimated pitches.
If the value of an entry is close to 1, we know that likely a music note with the fundamental frequency is active on that time frame.

First, we construct a \emph{harmonic map} that shifts the active entries in $\mathbf{P}_0$ upwards by a multiple of the corresponding fundamental frequency ($f_0$). That is, the $(t,f)$-th entry in the resulting harmonic map $\mathbf{P}_n \in \mathcal{R}^{258 \times 88}$ is nonzero only if that frequency is $(n+1)$ times larger than an active $f_0$ that frame, i.e., $f=f_0 \cdot (n+1)$.

Then, a harmonic series feature up to the $(n+1)$-th harmonics,\footnote{We note that the first harmonic is the fundamental frequency.} denoted as $\mathbf{H}_n  \in \mathcal{R}^{258 \times 88}$, is computed by an element-wise sum of $\mathbf{P}_0$, $\mathbf{P}_1$, $\dots$ up to $\mathbf{P}_n$, as illustrated in Figure \ref{fig: flow}. In that follows, we also refer to $\mathbf{H}_n$ as HSF--$n$.

When using HSF--$n$ as input to the instrument recognition model, we concatenate CQT $\mathbf{X}$ and $\mathbf{H}_n$ along the channel dimension, to the effect that emphasizing the partials in the input audio. The resulting matrix is then used as the input to a CNN model depicted in Figure \ref{fig: flow}. The CNN model used here is also adapted from \cite{SYnet}, using 1D convolutions, ResBlocks, and 11 convolutional layers in total. We call this model `\emph{CQT$+$HSF--$n$}' hereafter.



\subsubsection{Other Ways of Using Pitch} \label{subsubsec:pitch_implement}

We consider another two methods to use pitch information. 

First, instead of stressing the overtones, the matrix $\mathbf{P}_0$ already contains information regarding which pitches are active per time frame. This information can be important because different instruments (e.g., violin, viola and cello) have different pitch ranges. Therefore, a simple way of taking pitch information into account is to concatenate $\mathbf{P}_0$ with the input CQT $\mathbf{X}$ along the frequency dimension (which is fine since we use 1D convolutions), leading to a $258\times176$ matrix, and then feed it to the early convolutional layer. This method exploits pitch information right from the beginning of the feature learning process. We call it the `\emph{CQT$+$pitch~(F)}' method for short. 

Second, we can also concatenate $\mathbf{P}_0$ with the input CQT $\mathbf{X}$ along the channel dimension, to allow the pitch information to directly influence the input CQT $\mathbf{X}$. It can tell us the pitch note and onset timing, which is critical in instrument recognition. We call this method `\emph{CQT$+$pitch~(C)}'.

\begin{table*} [t]
\centering
\begin{tabular}{|c|l|ccccccc|c|}
\hline
Pitch & \multirow{2}{*}{Method} & \multirow{2}{*}{Piano}&\multirow{2}{*}{Violin}&\multirow{2}{*}{Viola}&\multirow{2}{*}{Cello}&\multirow{2}{*}{Clarinet}&\multirow{2}{*}{Bassoon}&\multirow{2}{*}{Horn}&\multirow{2}{*}{Avg.} \\ 
source &  & &&&&&&& \\ \hline\hline
\multirow{2}{*}{none} & CQT only (based on \cite{JYnet})&0.972&0.934&0.798&0.909&0.854&0.816&0.770&0.865\\ 
&CQT only (based on \cite{SYnet})&\textbf{0.982}&\textbf{0.956}&\textbf{0.830}&\textbf{0.933}&\textbf{0.894}&\textbf{0.822}&\textbf{0.789}&\textbf{0.887}\\ \hline
& CQT$+$HSF--1&\textbf{0.999}&\textbf{0.986}&\textbf{0.916}&\textbf{0.972}&\textbf{0.945}&0.909&0.776&0.929\\  
groud- & CQT$+$HSF--2&0.997&0.984&0.912&0.968&0.941&0.906&0.799&0.930\\  
truth & CQT$+$HSF--3&0.997&0.985&0.914&0.971&0.944&0.907&0.810&\textbf{0.933} \\  
pitch& CQT$+$HSF--4&0.997&\textbf{0.986}&0.909&0.969&0.944&0.904&\textbf{0.815}&0.932\\  
& CQT$+$HSF--5&0.998&0.975&0.902&0.968&0.942&\textbf{0.912}&0.803&0.928 \\
\hline
& CQT$+$HSF--1&0.983&0.955&\textbf{0.841}&0.935&0.901&0.822&0.793&0.890\\  
 & CQT$+$HSF--2&0.983&0.954&0.830&0.933&0.899&0.820&0.800&0.889\\  
estimated& CQT$+$HSF--3&0.983&0.955&0.829&0.934&0.903&0.818&\textbf{0.805}&0.890\\  
pitch& CQT$+$HSF--4&0.981&0.955&0.833&\textbf{0.937}&0.903&0.831&0.793&0.890\\  
by \cite{thickstun2018invariances}  & CQT$+$HSF--5&\textbf{0.984}&0.956&0.835&0.935&\textbf{0.915}&\textbf{0.839}&\textbf{0.805}&\textbf{0.896}\\
&CQT$+$Pitch~(F)&0.983&0.955&0.829&0.936&0.887&0.819&0.791&0.886\\ 
&CQT$+$Pitch~(C)&0.982&\textbf{0.958}&0.819&0.921&0.898&0.827&0.794&0.886\\ \hline 
\end{tabular}
\caption{Recognition accuracy (in F1-score) of model with and without pitch information, using either ground truth pitches or estimated pitches. We use bold font to highlight the best result per instrument for the three groups of results.}
\label{tab:result}
\end{table*}

\subsection{Implementation Details}
\label{subsec:experiment}

All the networks are trained using stochastic gradient descend (SGD) with momentum 0.9. The initial learning rate is set to 0.01. 
The weighted cross entropy, as defined below, is used as the cost function for model training:
\begin{equation} 
l_n = -y_n[t_n \cdot \log \sigma(\hat{y}_n)+(1-y_n)\cdot\log(1-\sigma(\hat{y}_n))] \,,
\label{eq:loss1}
\end{equation}
where 
$y_n$ and $\hat{y}_n$ are the ground truth and predicted label for the $n$-th instrument per time frame, $\sigma(\cdot)$ is the sigmoid function to reduce the scale of $\hat{y}_n$ to $[0,1]$, and $w_n$ is a weight computed to emphasize positive labels and counter class imbalance between the instruments, based on the trick proposed in \cite{googletrick}. 
Code and model are built with the deep learning framework PyTorch.

Due to the final sigmoid layer, the output of the instrument recognition model is a continuous value in $[0,1]$ for each instrument per frame, which can be interpreted as the likelihood of the presence for each instrument. To decide the existence of an instrument, we need to pick a threshold to binarize the result.
Simply setting the threshold to 0.5 equally for all the instruments may not work well. Accordingly, we implement a simple threshold picking algorithm that selects the threshold (from 0.01, 0.02, $\dots$ to 0.99, in total 99 candidates) per instrument by maximizing the F1-score on the training set.

F1-score is the harmonic mean of precision and recall. 
In our experiments, we compute the F1-score independently (by concatenating the result for all the segments) for each instrument 
and then report the average result across instruments as the performance metric. 

We do not implement any smoothing algorithm to postprocess the recognition result, though this may help \cite{liang14iclr}.

\begin{figure}[t]
\centering
\includegraphics[width=\linewidth]{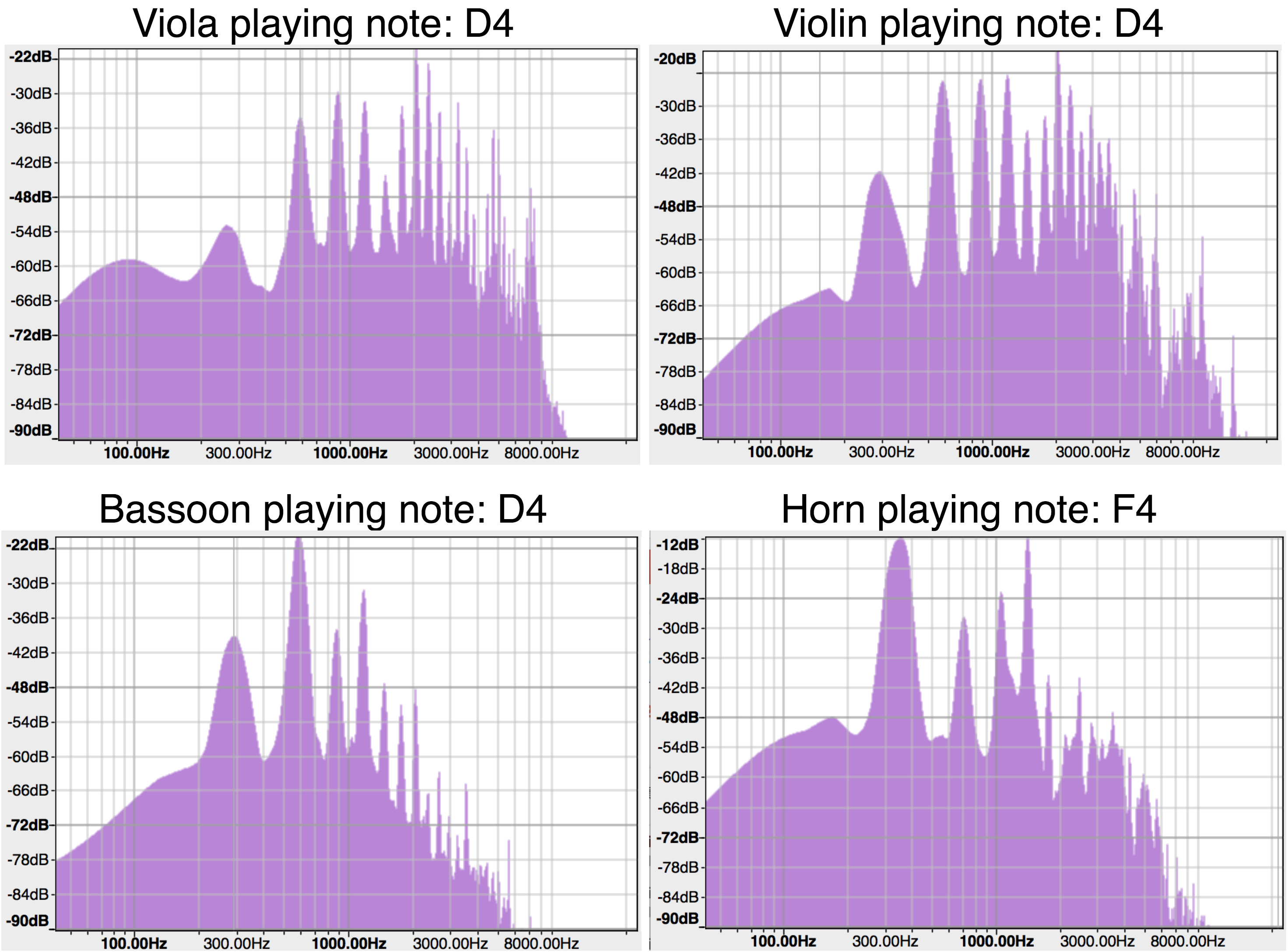}
\caption{Harmonic spectrum of Viola (top left), Violin (top right), Bassoon (bottom left) and Horn (bottom right), created by the software Audacity \cite{Audacity} for real-life recordings of instruments playing a single note.}
\label{fig:spectrum}
\end{figure}

\section{Performance Study} 
\label{sec:result}



The evaluation result is shown in Table \ref{tab:result}.
We first examine the result between two models without pitch information. From the first and second rows, we  see that adding Res-blocks indeed leads to a more accurate model. Therefore, we also use 
Res-blocks for the pitch-aware models.

We then examine the result when we use ground truth pitch labels to inform the model. From the upper half of Table \ref{tab:result}, pitch-aware models (i.e., CQT$+$HSF) indeed outperform the models that only use CQT. While the CQT-only model based on \cite{SYnet} attains 0.887 average F1-score, the best model CQT$+$HSF-3 reaches 0.933. Salient improvement is found for \emph{Viola}, \emph{Clarinet}, and \emph{Bassoon}. 

Moreover, a comparison among the pitch-aware models shows that different instruments seem to prefer different numbers of harmonics $n$. 
\emph{Horn} and \emph{Bassoon} achieve best F1-score with larger $n$ (i.e., using more partials), while \emph{Viola} and \emph{Cello} achieves best F1-score with smaller $n$ (using less partials). This is possibly because string instruments have similar amplitudes for the first five overtones, as Figure \ref{fig:spectrum} exemplifies. Therefore, when more overtones are emphasized, it may be hard for the model to detect those trivial difference, and this in turn causes confusion between similar string instruments. In contrast, there is salient difference in the amplitudes of the first five overtones for \emph{Horn} and \emph{Bassoon}, making HSF--5 effective.



\begin{figure*}[t]
\centering
\includegraphics[width=\linewidth]{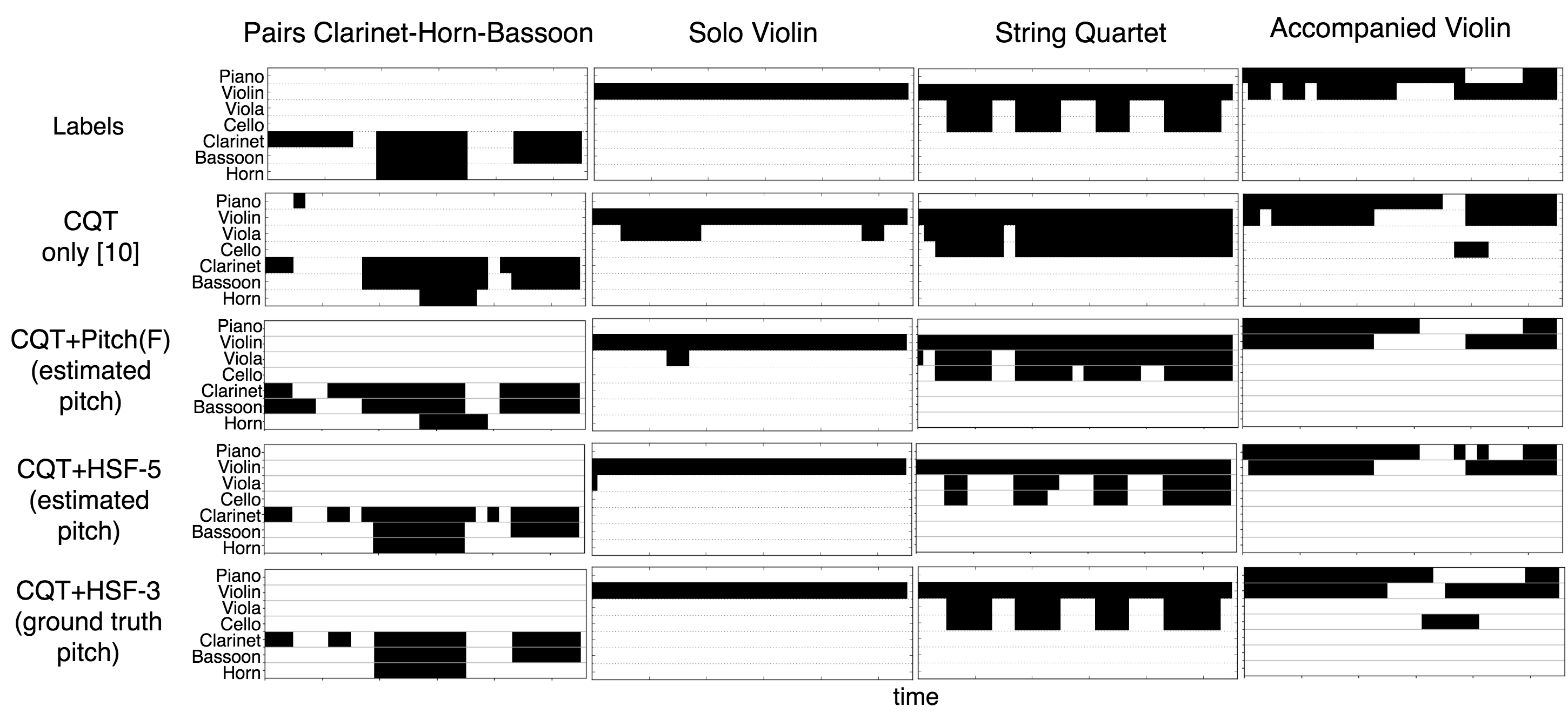}
\caption{Prediction results of different methods for four test clips. The first row shows the ground truth frame-level instrument labels, where the horizontal axis denotes time. The other rows show the frame-level instrument recognition result for a model that only uses CQT (`CQT only'; based on \cite{SYnet}) and three pitch-aware models that use either ground truth or estimated pitches. We use black shade to indicate the instrument(s) that are considered active in the labels or in the recognition result in each time frame.}
\label{fig:result}
\end{figure*}

Figure \ref{fig:result} shows qualitative result demonstrating the prediction result for four clips in the test set. By comparing the result of the first two rows and the last row, we see that onset frames are clearly identified by the HSF-based model. Furthermore, when adding HSF, it seems easier for a model to distinguish between similar instruments (e.g., violin versus viola). These examples show that adding HSF helps the model learn onset and timbre information.


Next, we examine the result when we use pitch estimation provided by the model of Thickstun \emph{et al.}
\cite{thickstun2018invariances}.
We know already from Table \ref{tab:dataset} that multi-pitch estimation is not perfect. Accordingly,  
as shown in the last part of Table \ref{tab:result}, the performance of the pitch-aware models degrades, though still better than the model without pitch information.
The best result is obtained by CQT$+$HSF--5, reaching 0.896 average F1-score. Except for \emph{Violin}, CQT$+$HSF--5 outperforms CQT-only for all the instruments. We see salient improvement for \emph{Viola}, \emph{Clarinet}, \emph{Bassoon} and \emph{Horn}, for which the CQT-only model performs relatively worse. This shows that HSF helps highlight differences in the spectral patterns of the instruments.

Besides, similar to the case when using ground truth pitch labels, when using the estimated pitches, we see that \emph{Viola} still prefers using fewer harmonic maps, whereas \emph{Bassoon} and \emph{Horn} prefer more. Given the observation that different instruments prefer different number of harmonics, it may be interesting to design an automatic way to dynamically decide the number of harmonic maps per frame, to further improve the result.



\begin{figure}[t]
\centering
\includegraphics[width=\linewidth]{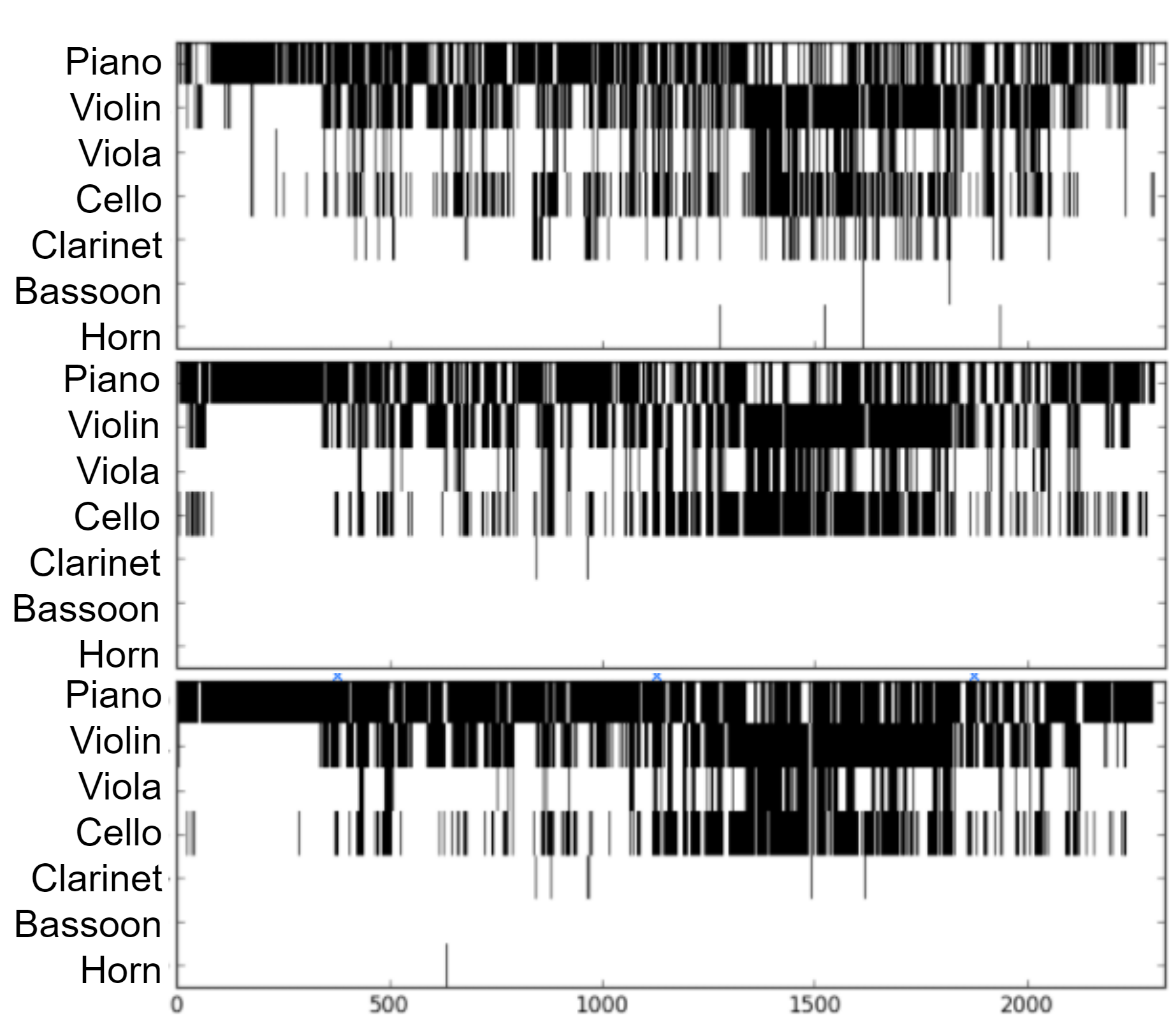}
\caption{Frame-level instrument recognition result for a pop song, \emph{Make You Feel My Love} by Adele, using the baseline CNN \cite{JYnet} (top), CNN + Res-blocks \cite{SYnet} (middle) and CQT$+$HSF--5 using estimated pitches (bottom).}
\label{fig:test}
\end{figure}

The fourth row of Figure \ref{fig:result} gives some result for CQT$+$ HSF--5 based on estimated pitches.
Compared to the result of CQT only (second row), we see that CQT$+$HSF--5 nicely reduces the confusion between \emph{Violin} and \emph{Viola} for the solo violin piece, and reinforces the onset timing for the string quartet piece.

Moving forward, we examine the result of the other two pitch-based methods, CQT+Pitch~(F) and CQT+Pitch~(C), using again estimated pitches. 
From the last two rows of Table \ref{tab:result}, we see that 
these two methods do not perform better than even the second CQT-only baseline. 
As these two pitch-based methods take the pitch estimates directly as the model input, we conjecture that they are more sensitive to errors in multi-pitch estimation and accordingly cannot perform well. From the recognition result of the string quartet clip in the third row of Figure \ref{fig:result}, we see that the CQT+Pitch~(F) method cannot distinguish between similar instruments such as \emph{Violin} and \emph{Viola}. 
This suggests that HSF might be a better way to exploit pitch information.

Finally, out of curiosity, we test our models on a famous pop music (despite that our models are trained on classical music). 
Figure \ref{fig:test} shows the prediction result for the song \emph{Make You Feel My Love} by Adele. It is encouraging to see that our models correctly detect the \emph{Piano} used throughout the song and the string instruments used in the middle solo part. Moreover, they correctly give almost zero estimate for the wind and brass instruments. Moreover, when using the Res-blocks, the prediction errors on \emph{clarinet} are reduced. When using the pitch-aware model, the prediction errors on \emph{Violin} and \emph{Cello} at the beginning of the song are reduced. Besides, \emph{Piano} timbre can also be strengthened when \emph{Piano} and the strings play together at the bridge.  


\section{Conclusion}\label{sec:conclusion}
In this paper, we have proposed 
several methods for frame-level instrument recognition. Using CQT as the input feature, our model can achieve 88.7\% average F1-score for recognizing seven instruments in the MusicNet dataset. Even better result can be obtained by the proposed pitch-aware models. Among the proposed methods, the HSF-based models achieve the best result, with average F1-score 89.6\% and 93.3\% respectively when using estimated and ground truth pitch information.

In future work, we will include MedleyDB to our training set to cover more instruments and music genres. We also like to explore joint learning frameworks and recurrent models (e.g., \cite{CRNN-tagging,LSTM-tagging,OnsetAndFrame}) for better accuracy.


\section{Acknowledgement}\label{sec:acknowledgement}
This work was funded by a project with KKBOX Inc.

\bibliography{ref}
\end{document}